\documentclass[letterpaper,twocolumn,10pt]{article}
\usepackage{xcolor}  
\usepackage[colorinlistoftodos,prependcaption,textsize=tiny,disable]{todonotes}%
\usepackage{xargs}                      
\usepackage{mdframed}
\usepackage{ntheorem}

\newif\ifistoreview

\istoreviewfalse

\theoremstyle{nonumberplain}
\newmdtheoremenv[%
  linecolor=blue,
  linewidth=2pt,
  rightline=false,
  leftline=false]{figrev}{}

\graphicspath{{figures/}}

\begin{document}

\title{\vspace*{-1cm} Flexibility Is Key in Organizing a Global Professional Conference Online: The ICPE 2020 Experience in the COVID-19 Era \vspace*{0.5cm}}

\author{
 {\rm Alexandru Iosup} \\ VU Amsterdam, \\ The Netherlands \\ A.Iosup@vu.nl
 \and  {\rm Catia Trubiani} \\ Gran Sasso Science Institute, \\ 
Italy \\ catia.trubiani@gssi.it 
 \and  {\rm Anne Koziolek} \\ KIT, \\ Germany \\ koziolek@kit.edu
 \and  {\rm Jos\'{e} Nelson Amaral} \\ University of Alberta, 
\\ Canada \\ jamaral@ualberta.ca 
 \and  {\rm Andre B. Bondi} \\ Software Performance and \\ Scalability Consulting LLC, USA \\ AndreBBondi@gmail.com
 \and  {\rm Andreas Brunnert} \\ RETIT GmbH, \\ Germany \\ brunnert@retit.de
 }

\maketitle

\begin{abstract}
Organizing professional conferences online has never been more timely. 
Responding to the new challenges raised by COVID-19, 
the organizers of the ACM/SPEC International Conference on Performance Engineering 2020 had to address the question: How should we organize these conferences online? 
This article summarizes their successful answer.
\end{abstract}

\section{Introduction}
\label{sec:introduction}\label{sec:intro}

The evolving travel and gathering restrictions caused by the COVID-19 pandemic created a climate of uncertainty that made for difficult planning and decision making about international conferences scheduled for April 2020. Tribute to the ingenuity of our field, a new model quickly evolved, pioneered by conferences such as ASPLOS~\cite{blog:cacm:LarusCS20} and EDBT/ICDT~\cite{blog:cacm:Bonifati+20}, and by ACM itself~\cite{report:acm20}. This article reports on a design alternative to organizing professional conferences online, derived from our experience with organizing the 11th ACM/SPEC International Conference on Performance Engineering (ICPE 2020).

Particular to the ICPE timing was the process to decide about holding an in-person conference, postponing to a later date, or converting the conference to an online event. As late as mid-February, we were unaware of the extent of the spread of COVID-19 in Canada, where the conference was due to occur, or in the countries from which our attendees were expected to travel. This was not due to our ignoring the news, but to the deficit of credible information offered by state authorities. On March 9, one of the Program Chairs received the final institutional decision of being stopped from international travel; one day later, US- and EU-based keynote speakers for workshops started to declare their unavailability. Discussions with relevant stakeholders -- local organizers, local businesses and Horeca, international organizations -- started, as did the consultation of relevant official guidelines\footnote{Advice from the Canadian Government:  \url{https://www.canada.ca/en/public-health/services/diseases/2019-novel-coronavirus-infection/health-professionals/mass-gatherings-risk-assesment.html}}.
On March 13, we announced {\it “ICPE 2020 will not be held in April 2020 as planned”}. On March 19, we discussed in the Steering Committee of ICPE two options regarding the physical meeting: cancelling and rescheduling for what we predicted as a safer period, late-July\footnote{To arrive at the month of July, we asked an in-house statistician to predict the period expected to exhibit a lowering of COVID-19 presence. The statistician used the SIR model on WHO data of SARS cases. See David Smith and Lang Moore, "The SIR Model for Spread of Disease - The Differential Equation Model," Convergence (December 2004) and the WHO Epidemic curves - Severe Acute Respiratory Syndrome (SARS) Nov 2002 to Jul 2003: \url{https://www.who.int/csr/sars/epicurve/epiindex/en/index1.html}
}.

We discarded through discussion the idea of re-inviting the authors of accepted articles for next year’s conference. We decided to opt for a cancellation of the physical conference, which was in line with our moral choice to protect the community and was also supported by our main technical sponsors. We also formed a Task Force for Organizing ICPE Online. On April 2, following extensive documentation, discussions, and try-outs, the Task Force decided to propose organizing the 11th ICPE online. We were strongly motivated by what we saw as our duty to deliver to the community, but also by our belief that we could do so through an online event. We saw it as our duty to “give justice” to the authors of accepted papers and to guarantee that they can get the same level of feedback they would have received if attending the physical meeting, and the online option seemed to give us an opportunity to achieve this duty.

The 11th ICPE was held fully online, on April 20-24, 2020, with two days of workshops and three days of single-track conference sessions. Key to our approach to organizing ICPE online was to be flexible by design, mixing synchronous and asynchronous events, and giving the attendees many options to participate and contribute, while ensuring all of the original sessions of the conference maintained a synchronous element. We also ensured free registration and free publication of proceedings, which were only possible thanks to our generous supporters and sponsors. These two elements, and others, distinguish our design for the conference from previously reported designs~\cite{blog:cacm:LarusCS20,blog:cacm:Bonifati+20}. In return, we observed a significant increase in registered participants, and both the synchronous and the asynchronous channels were very well attended, exceeding the audience of the physical ICPE meetings in the past couple of years.

In hindsight, many of our decisions, starting with organizing ICPE 2020 online, seem obvious. However, this was not the case at the time, and luckily by the end of the process many were satisfied. To quote one of the senior members of the community: {\it ‘I was one of the sceptics for an online ICPE (or any other online conference for that matter). But you really proved me wrong :) It was a great event’}. There were only 3 weeks to organize everything, and we greatly appreciated the existing guidelines and reports (\cite{report:acm20,blog:cacm:LarusCS20} and, from April 8, also~\cite{blog:cacm:Bonifati+20}). For these reasons, we present here a summary of our design choices, the experience of the online ICPE 2020, the feedback collected from the community, and the lessons we learned as organizers.

The remainder of this document is structured as follows.
Section~\ref{sec:design} presents our design choices for ICPE 2020.
Section~\ref{sec:execution} summarizes the execution of the online conference.
Section~\ref{sec:feedback} reports on community feedback; further feedback appears in the Appendix.
Last, Section~\ref{sec:conclusion} concludes and summarizes our advice for organizers of future conferences.


\section{Design Choices for ICPE 2020} \label{sec:design}

The reliability of communication over the Internet is an important consideration when organizing a meeting 
online~\cite{report:acm20,blog:cacm:LarusCS20,blog:cacm:Bonifati+20}. However, based on our past experience with online communities and especially gaming, and now also with ICPE, we believe Internet reliability is not the key issue for organizing a professional meeting between motivated participants. It is probably a confirmation bias: in hindsight, in most cases the online operation of ICPE was not more problematic than of an on-site (physical) operation\footnote{Some of the failures that occur when organizing on-site include microphone not working, speaker not having a compatible interface for the video projector, speaker trying to use pointers on a mispositioned screen, video projector failing or not displaying colors correctly, room too dusty or drafty, signaling difficult to follow, etc.}.

Our main insight is that {\bf the most important factor for the success of a professional meeting is the human factor}. Both {\it availability} and {\it attention}, which for online conferences can be limited by time differences and a variety of factors, are thus the key problems in need of good solutions. Also important, but less in the hand of the organizers and more a consequence of the community itself, is to have an interesting program and attractive talks. Thus, we set as our key principle to {\bf be as flexible as possible}, incentivizing people to attend and facilitating their interactions. 

We present in the following eight design choices. 
How they were perceived by the community will become clear in Section~\ref{sec:feedback}.

\subsubsection*{Q1: How to share the conference material, flexibly?}

{\bf A1}: At a minimum, a conference needs to allow its participants to access the written material (the publications) and to jointly see the presentations. We knew the ACM proceedings would be available at the start of the main conference and would be free to access~\cite{proc:icpe20,proc:icpe20ws}, but this precluded them from being seen earlier, and in particular did not allow the workshop attendees to access them---because the workshops were scheduled for the two days before the start of the main conference. We couldn’t publish the articles internally, due to copyright issues. Our solution was to ask the authors to {\it publish pre-prints}. Guidelines, email communication, archival offices and arxiv.org, and good will were necessary to achieve this. We also asked authors to create {\it videos of their talks} the moment we decided to cancel the physical meeting, and share these through the ICPE YouTube channel~\cite{youtube:icpe20}; this decision would become useful in a later design decision. After deciding to go online, we further asked the authors to create and share {\it 1-slide pitches} of their work, and {\it full slide-decks} explaining their work. We not only linked to these on the ICPE website~\cite{website:icpe20}, but also asked the authors to share links to their papers via the ICPE Slack workspace~\cite{slack:icpe} and also on social media (Twitter and LinkedIn, primarily). This abundance of material and channels put additional burden on the authors, but allowed the attendees the flexibility of choosing how to consume the information.

\subsubsection*{Q2: How to facilitate attendance for all possible members of the community?}

{\bf A2}: We realized early that attendance is significantly limited by any financial burden. Thanks to generous funding from both technical sponsors (the ACM SIGmetrics and the ACM SIGSoft), from the SPEC Board of Directors, and from Huawei (generously confirming their Silver Sponsorship), and Samsung, we were able to reimburse the author’s registrations in full and to make the proceedings freely accessible online. With the help of the Faculty of Science at the University of Alberta, and through a new sponsorship from the SPEC Board of Directors, we were then also able to offer full, {\it free access to the ICPE 2020 events}. As we will see, this important decision allowed students and other first-time participants to attend, without a financial burden.

\subsubsection*{Q3: Which infrastructure to use for organizing the conference online?}

{\bf A3}: We decided to use a set of largely complementary software tools to build the infrastructure for the conference. Flexibility and ease of use were prime considerations, and we ended up using what we think are the best tools currently available for the job: email and website for one-way, asynchronous communication; Zoom (primary) and GoToMeeting (secondary) for multi-party, synchronous video communication and (to some extent) online messaging; YouTube for one-way, asynchronous video communication; Google Drive (primary) and various archival services (also primary) for asynchronous one-way sharing of files; Slack for multi-party, synchronous and asynchronous messaging and file-sharing (secondary); and TimeAndDate to share official times in the format needed by the attendees. We also set up a Mozilla Hubs breakout room -- a virtual space displaying in the browser or on a VR headset -- but the audience did not try it much. (We have also considered many alternatives, especially for Zoom.) This allowed us to support various modes of communication, much like a well-equipped physical facility would allow; it also required us the organizers to act as the technical team in ways that would normally be delegated to the manager of the physical facility.

\subsubsection*{Q4: How to organize the sessions, to maximize availability and attention?}

{\bf A4}: We addressed this genuinely difficult question, which also appears to be the crux of education and training in general, through a set of measures. We describe here a selection of such measures:

\begin{enumerate}
    \item We aimed to improve attention, at the possible cost of availability, by organizing all the events of the conference synchronously, that is, all the keynotes, talks, and moderated Q\&A parts are attended by the audience when they occur. This decision was also taken by EDBT/ICDT, but contrasts to the asynchronous organization of ASPLOS [2].

    \item We aimed to improve both availability and attention by limiting the duration of the virtual “day” to about 3 hours, aligned with the time-zone of the original location of the conference (5pm in Central Europe is 9am in Edmonton, Canada). This limited somewhat potential participation from Asia, but time-zones are very limiting and we reasoned the material is available online in any case. This also limited the duration we can allocate per article, for example, but in our view allowed the attendees to still have enough energy to engage beyond the session itself -- a behavior we have observed is that both speakers and authors would continue to engage on Slack, synchronously or asynchronously.

    \item To improve attention, we further limited the duration of each keynote to 25 minutes of one-way communication and 5+ minutes of Q\&A, and asked presenters of peer-reviewed articles to stay within a budget of about 20 minutes for full articles and 15 minutes for short articles (more about this in Q5).

    \item We aimed to improve availability, by making all the material available for offline access. This includes not only the material from Q1, but also the Q\&A and discussions. For the latter, we arranged to have helpers from the community (one secretary per session and other self-appointed participants) transcribe to the Slack. We noticed this decision has benefited attendees who could not be present, due to other commitments or illness.

    \item We aimed to improve attention, by asking for each session a moderator plus a small team selected from PC members to revise slides and videos, and to write on Slack questions for each article presented in the session, prior to the session itself. This focused the community’s attention, and also removed the obstacle of asking the first question.

    \item To improve both availability and attention, we aimed to select only tools we considered easy-of-use and appropriate for how conferences work. This led us to, in the end, reject the use of the Webinar mode of Slack and GoToMeeting---the Q\&A sessions were confusing, with participants not able to understand immediately how to ask questions or whether questions have been asked at all. Furthermore, tools like Slack offer many options and proved to be confusing for some in the audience.  

    \item To improve availability, we opted to have backup infrastructure for any mode of communication, e.g., Zoom and GoToMeeting for multi-party, synchronous video communication; YouTube and Slack for sharing videos; and Google Drive and Slack for various files. This led to higher costs (e.g., due to licenses), but gave us certainty the event could proceed even if one of the software tools suffered a catastrophic outage. This would be difficult to replicate with a physical organization of the conference.

    \item To improve availability, we opted for Zoom as the primary infrastructure for multi-party, synchronous video communication, coupled with Slack as primary tool for online messaging. In our experience, among at least five other leading platforms, Zoom worked best, being easy to install and use, and exhibiting very few hiccups even when the attendance scaled or became global. (We will not comment on these features, or on the security and privacy issues related to Zoom and other platforms. We are not aware of an impartial, high-quality, reproducible study across all these platforms. Perhaps this is a topic for ICPE 2021...)

\end{enumerate}

\subsubsection*{Q5: How to organize the “talks”: keynotes, article-presentations, etc.?}

{\bf A5}: Keynotes are typically the star of general participation, so we decided to preserve their typical organization as one-way talks followed by a moderated Q\&A session. However, for talks related to peer-reviewed articles we reasoned that sitting through one-way communication would diminish the energy in the “room”. Thus, we diverged from the classic organization of conferences, and (1) asked the presenters to pre-record and share their talks on YouTube with at least several days before the first day of the main conference, (2) assigned the moderator and a team of experts to view the videos and ask questions prior to the “live” session, (3) encouraged the audience to do the same, (4) asked the speakers to pitch their work for up to 2 minutes at the start of the session, and (5) enabled and encouraged “live” questions, which were asked either by attendees live or, when technical glitches or personal preference precluded this, by the moderator. This flexible approach led to numerous questions and a lively discussion, perhaps even more than some articles would see in conferences organized classically.

\subsubsection*{Q6: How to ensure that everyone knows what to do?}

{\bf A6}: Communication is key in any process of change, and it was also in our case. We used every channel at our disposal to communicate with the audience, first through email, then through an extensive booklet with guidelines, last but not least through online meetings and Slack. The booklet, “Guidelines for the virtual ICPE 2020”, went through 3 major versions (uploaded on Slack), and on only 6 pages described the key terms of the meeting, pointed out the Code of Conduct\footnote{https://icpe2020.spec.org/code-of-conduct/}, and informed various personas (e.g., authors, session chairs, other attendees) about how to easily join the sessions. We clarified many aspects using Slack and, especially among the organizers and the Task Force, through online meetings.

\subsubsection*{Q7: How to facilitate the organization of ICPE-related events, flexibly?}

{\bf A7}: We again put the guiding principle of flexibility into practice: we offered advice and guidelines to organizers of workshops, but in the end were supportive with any choice they made. For example, one of the workshops decided to maintain the classic approach, with long keynotes and talks leading to about 8 hours in the virtual “day”; midway through the event, the European participants had to leave, because it was already late in their evening. 

\subsubsection*{Q8: How to facilitate social events, flexibly?}

{\bf A8}: We discussed extensively whether we should try to organize social events for the conference. In the spirit of flexibility, the answer can only be: let the society decide itself! And so we did. From the first days of the conference, it emerged that at the end of each day a sizable part of the community would simply “hang out” online, some with drinks, some chatting, some simply staying online. We also noticed that groups would “go to Slack”, continuing the discussion, as reflected by the written messages. Last, we suggested that Slack could also be the host of “private groups”, where attendees with similar interests (and, as it turns out, also attendees with similar background) could meet and arrange further messaging or even new Zoom or other video communication.


\section{Executing the Online ICPE 2020}
\label{sec:execution}

Executing ICPE 2020 was challenging, but rewarding. We describe the following setup and several observations.

\subsubsection*{Overall and Daily Setup}

We asked attendees to register to the conference (for free), and invited all registrants (nearly 550) to the ICPE 2020’s Slack workspace. From these, over 480 accepted the invitation and became attendees (see also Observation O1).  We announced a new Zoom link each day; with the link used by all sessions, which allowed all attendees to find a way to join (if they wanted to).

Our Slack setup was similar to that of ASPLOS, with channels for: the introductory session (1 channel), each keynote (2x), the awards session (1x), each session of the main conference (7x), and posters and demos (1x). Overall, each of these channels was well attended (see also O2). We also created the icpe-2020 channel for chair announcements, general for anyone to share, all-sessions for notifications about sessions by session moderators, random for informal conversations, support for asking for help. We also created private channels for organizers (org for the conference chairs, org-ws for the workshops chairs). Overall, we observed that each topical channel was well-attended, but the more general channels were too numerous and generated confusion.

Last, we saw the community create new channels, both public channels for emerging topics (listed as topic-x to group in the Slack interface) and private channels (many organized by people from the same geographical area).

When using Slack, we found that it is important to carefully edit permissions and settings. Slack is aimed at teams and thus has less restrictive default settings. For example, we learned late that organizers should disable displaying email addresses in member profiles and disable that members can use @channel to notify all other members in larger conferences to avoid unsolicited advertisement.

All moving parts considered, the execution was relatively uneventful. The organizers were on-site and addressed the occasional technical glitches (e.g., Zoom crashed), presenter issues (e.g., not sharing the right screen), etc.

\subsubsection*{O1: ICPE 2020 had unusually high attendance.}

\begin{figure}[!t]
    \centering
    \includegraphics[width=\linewidth]{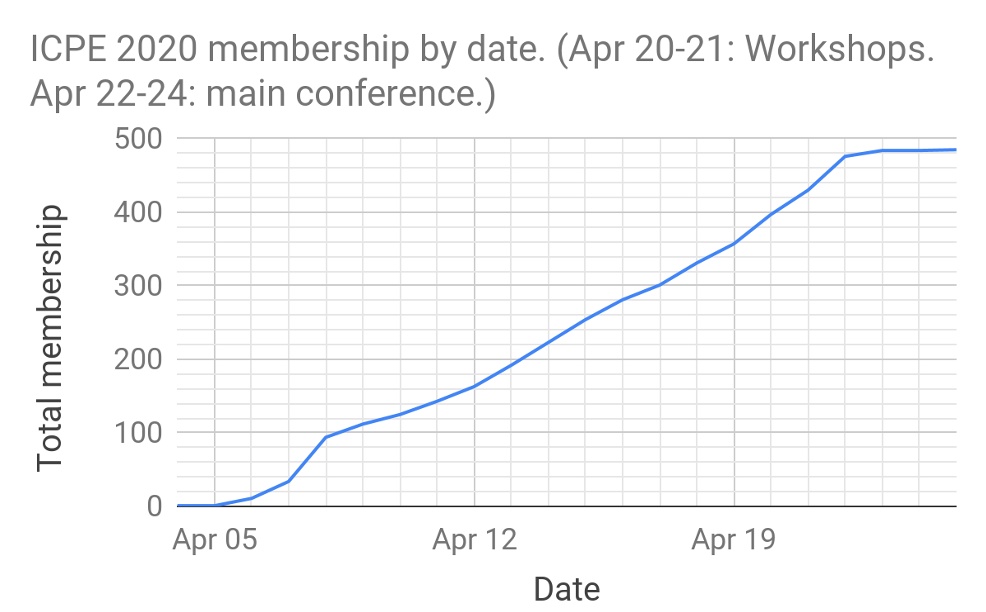}
    \caption{ICPE 2020 attendee count, by date. (Data until and including Apr 25.)}
    \label{fig:membership}
\end{figure}

Figure~\ref{fig:membership} depicts the growth in daily attendee count. We exceeded the expected number of attendees (150) by April 12, about 1 week after opening the registration process. At the start of the main conference, we already had over 480 attendees, which triggered us to upgrade the Zoom account to allow for more concurrent seats. This number of attendees is the highest ever recorded for ICPE, exceeding among others the participation observed in the previous edition, an extremely successful event organized in India. 

\subsubsection*{O2: Testing every aspect of the infrastructure is vital.}

\begin{figure}[!t]
    \centering
    \includegraphics[width=0.75\linewidth]{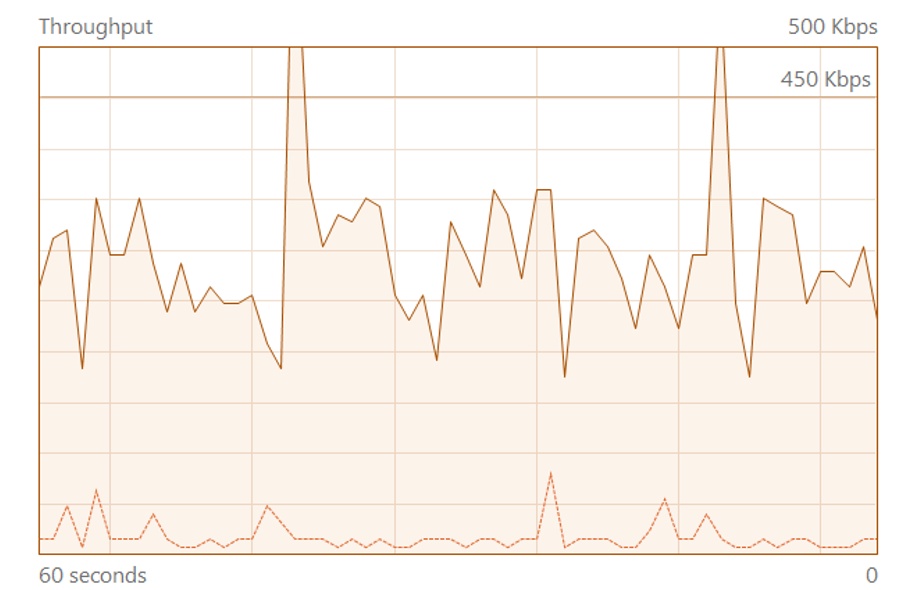}
    \includegraphics[width=0.20\linewidth]{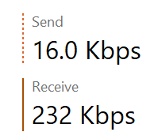}
    \caption{Bandwidth requirements for Zoom. (Live session with 1 main speaker, 1 moderator, circa 70 attendees.)}
    \label{fig:bw}
\end{figure}

We tested the infrastructure extensively, including during the event. Figure~\ref{fig:bw} depicts a bandwidth measurement conducted during the event (“load testing”), indicating the performance of Zoom at the scale needed by the conference remains relatively stable and affordable for reasonable Internet connections.

\subsubsection*{O3: The conference continues long after its last session.}

A conference does not have to end with its last session. We conducted surveys to obtain feedback from the attendees, compiled a report for the Steering Committee (this report), and started work on next year’s conference. But two items deserve more discussion: 

\begin{itemize}

    \item Although Slack preserves the written conversation, its forum capabilities are limited, so discussions may quickly become difficult to traverse. Following a tradition that exists in performance engineering since at least the late 1950s~\cite{DBLP:conf/ifip/Strachey59}, we have decided to summarize all Q\&A sessions, by item of discussion, through a community effort. The process is ongoing.
    
    \item We observed that the community has put effort into  identifying four emerging topics, which could lead to new entries in future ICPE conferences: {\tt topic-datasets}, about sharing datasets in the community, {\tt topic-edu}, about creating an education workshop associated with ICPE, {\tt topic-history}, about writing a (short) history of the field, and {\tt topic-per-var}, about understanding and controlling performance variability in software and hardware systems.
    
\end{itemize}

\subsubsection*{O4: Activity is most intense during the conference.}
The live participation was lower than the maximum possible, but still exceeded our expectation. On Zoom at peak, we counted over 175 concurrent attendees for the main conference, and 70-90 participants for the workshops. The lowest attendance was around 70 participants in the main conference, and 40-60 for the workshops. (We could not access the Zoom statistics for these metrics, so we counted them manually.)

\begin{figure*}[!t]
    \centering
    \includegraphics[width=\linewidth]{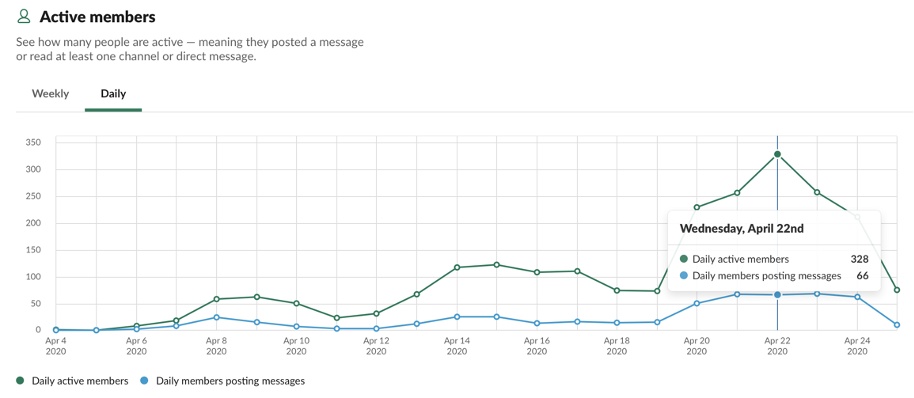}
    \caption{Number of daily active members and of daily active members posting messages.}
    \label{fig:activemembers}
\end{figure*}

Figure~\ref{fig:activemembers}, which depicts the number of daily active members and of daily active members posting messages, leads to our observation that activity is most intense only during the conference. On Slack, we counted over 325 daily active members. The peak is recorded during the opening of the conference, when in particular the SPEC community attended (increase in the daily active members) but not necessarily engaged in the discussion (similar count of members posting messages as in the previous days). This indicates that more community management and engagement is needed, to make such a community thrive beyond the limits of the ICPE event.

\subsubsection*{O5: The workshops are important contributors to the discussion.}

\begin{figure*}[!t]
    \centering
    \includegraphics[width=\linewidth]{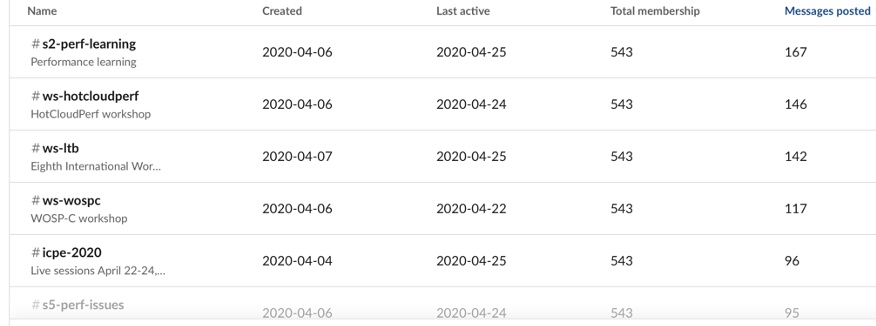}
    \caption{Number of messages posted in the public channels associated with ICPE main-conference sessions and workshops. Only the Top-6 of the 30 public channels displayed. (Data until Apr 25.)}
    \label{fig:messages}
\end{figure*}

We often hear the argument that workshops bring into the community hot topics of discussion, and overall can make conferences livelier. Figure~\ref{fig:messages} presents quantitative evidence in this sense: from the Slack channels dedicated to each session, the workshops stand out as 3 of the Top-5 sessions with the largest message count. (Public channels also accounted for 85\% of the message-views, so their impact of workshops was high.)

%
\newpage{}
\ \\
\newpage{}
\ \\
\newpage{}
\section{Community Feedback}
\label{sec:feedback}

We have conducted a comprehensive survey with over 50 questions, which we analyzed when it reached 50 respondents (just over 10\% of the attendees, and over one-quarter of the peak of concurrent attendees on Zoom). The participation was diverse in role in ICPE 2020 (about 40\% were authors, about a quarter speakers), current occupation (PhD students, academic staff, and industry engineers each represent over 20\% of the respondents), seniority (about half were seniors with 15+ years of experience, about 25\% were juniors), ownership of a PhD degree (about half), geolocation (about half from Europe and 40\% North America, with over 10 countries represented in the survey), and gender (one-third not male). We present here a selection of the results, with more results in the Appendix.

\begin{figure}[h]
    \centering
    \includegraphics[width=\linewidth]{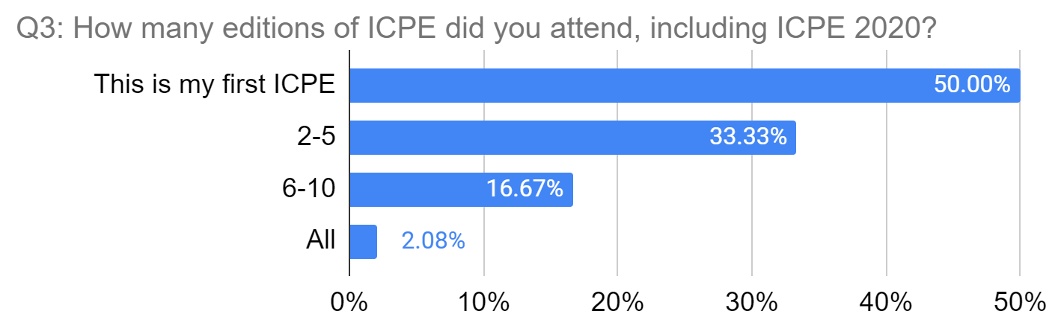}
    \caption{Summary of answers for Q3.}
    \label{fig:Q3}
\end{figure}

From Q3~(Figure~\ref{fig:Q3}), we observe that organizing online helped us enlarge participation, with about half of the respondents being first-time attendees.

\begin{figure}[h]
    \centering
    \includegraphics[width=\linewidth]{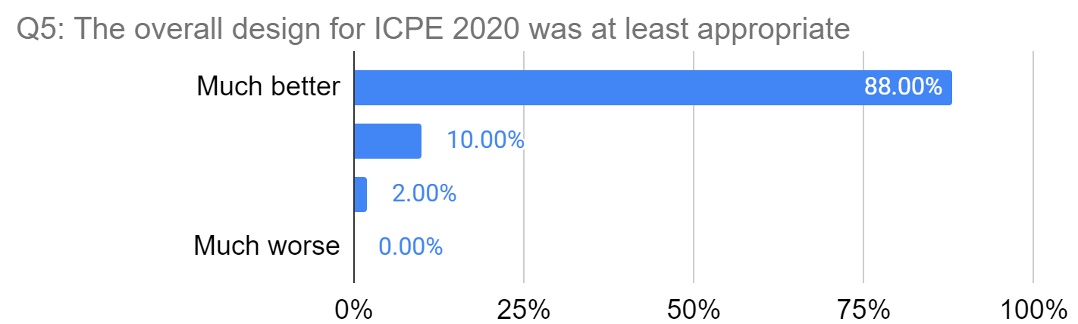}
    \includegraphics[width=\linewidth]{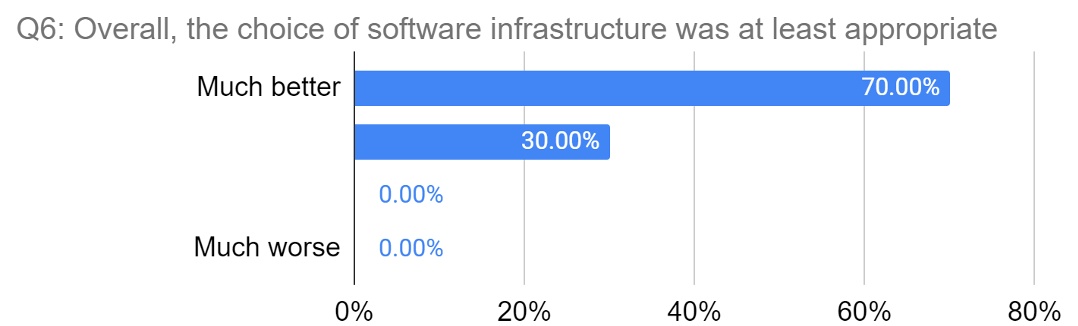}
    \caption{Summary of answers for Q5 and Q6.}
    \label{fig:Q56}
\end{figure}

From Q5 and Q6~(Figure~\ref{fig:Q56}), we observe that the attendees appreciated the organization, both in terms of design choices and in which software infrastructure was selected. We will see later that the audience did not like some of the features of an online conference as much. From the foregoing, we conclude that there is a need for new software that better supports the type of online conference we ran. On the other hand, the strong attendance and favourable feedback at the end of the conference indicate that we made successful use of the tools available, despite intense time pressure.

\begin{figure}[h]
    \centering
    \includegraphics[width=\linewidth]{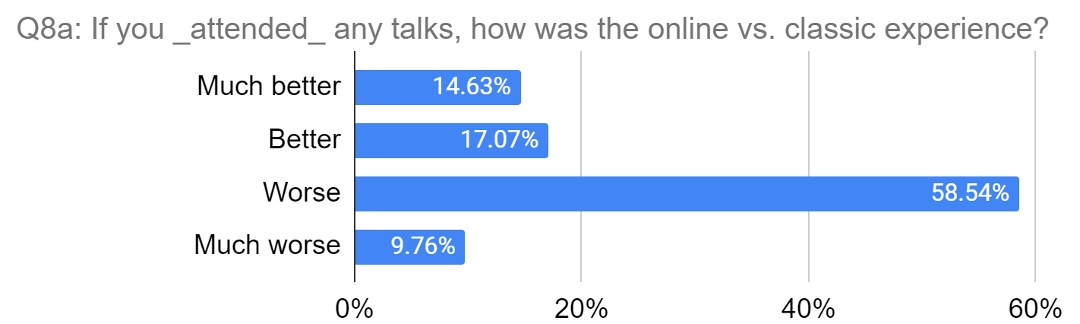}
    \includegraphics[width=\linewidth]{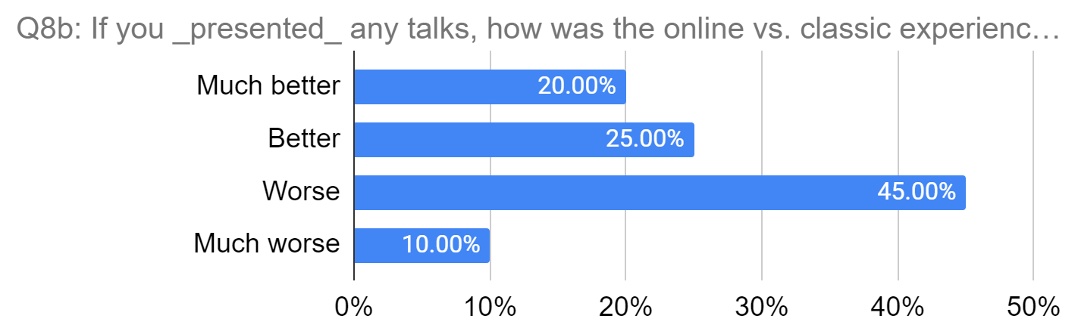}
    \includegraphics[width=\linewidth]{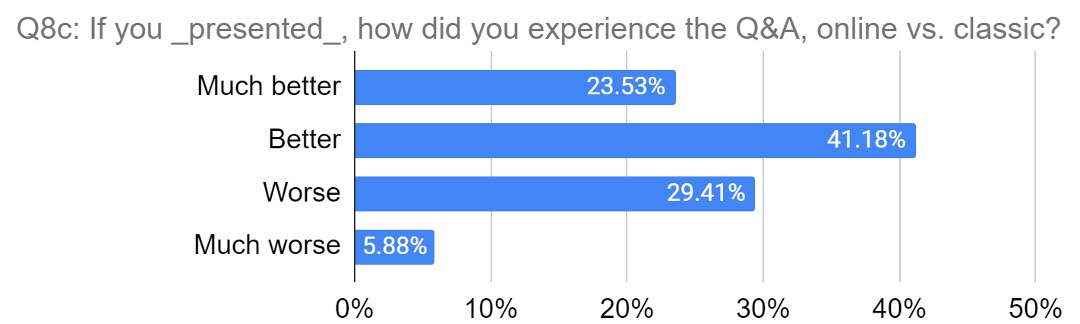}
    \caption{Summary of answers for Q8.}
    \label{fig:Q8}
\end{figure}

Overall from Q8~(Figure~\ref{fig:Q8}), we observe that the attendees experienced the online talks as worse for the online ICPE than for the classic. There was also enough appreciation, with between one-third and just less than half of the attendees appreciating the online approach more. This was consistent across both talks and keynotes, and both for speakers and audience. This matches the findings of ASPLOS and EDBT/ICDT, if we assume most of their borderline decisions would be accounted to our “Worse” category.

The results for the moderated Q\&A sessions were much appreciated. About two-thirds of the attendees, both speakers and audience, considered the online event at least better, and about one-quarter considered it much better.  This confirms our own observation that there was much more extensive and deeper interaction than we observed in the conventional format. The Slack channels were buzzing long after the session ended.

\begin{figure}[h]
    \centering
    \includegraphics[width=\linewidth]{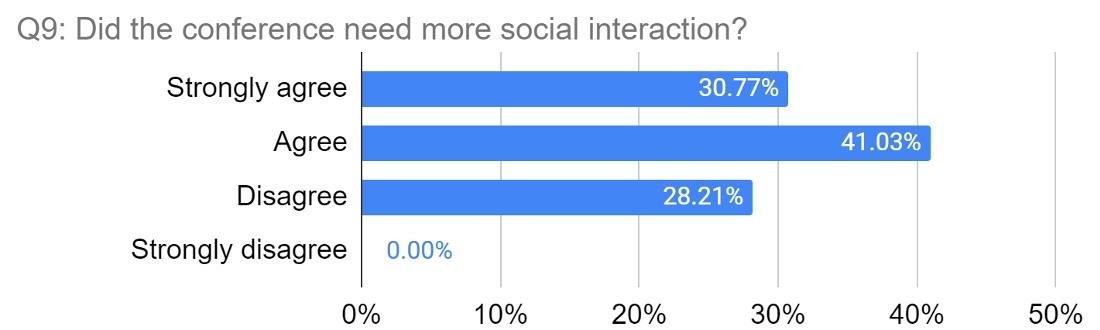}
    \caption{Summary of answers for Q9.}
    \label{fig:Q9}
\end{figure}

From Q9~(Figure~\ref{fig:Q9}), we conclude the need for social interaction remained unfulfilled. Over two-thirds of the respondents would have preferred more such interaction. This is consistent with the findings of the EDBT/ICDT survey. In our view, better tools (and maybe also better format-designs) need to appear before the online community can be fully satisfied about this aspect.

\begin{figure}[h]
    \centering
    \includegraphics[width=\linewidth]{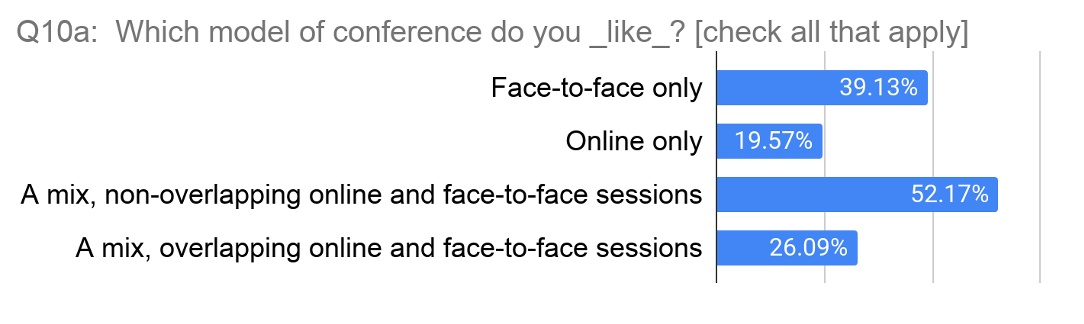}
    \includegraphics[width=\linewidth]{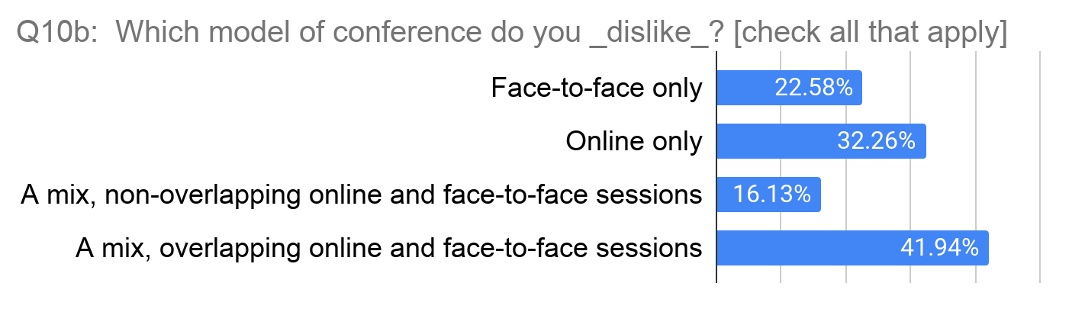}
    \caption{Summary of answers for Q10.}
    \label{fig:Q10}
\end{figure}

From the answers to Q10a and Q10b~(Figure~\ref{fig:Q10}), we observe that: (1) as one might expect, many would like a face-to-face only conference; (2) surprisingly, most would enjoy attending a mix of online and face-to-face sessions (but would prefer these sessions do not overlap); (3) a surprisingly high fraction of respondents, about one-fifth, dislike face-to-face conferences (but attended an online form!); (4) perhaps unsurprisingly, about one-third of the respondents dislike the idea of online only conferences.



\begin{figure}[h]
    \centering
    \includegraphics[width=\linewidth]{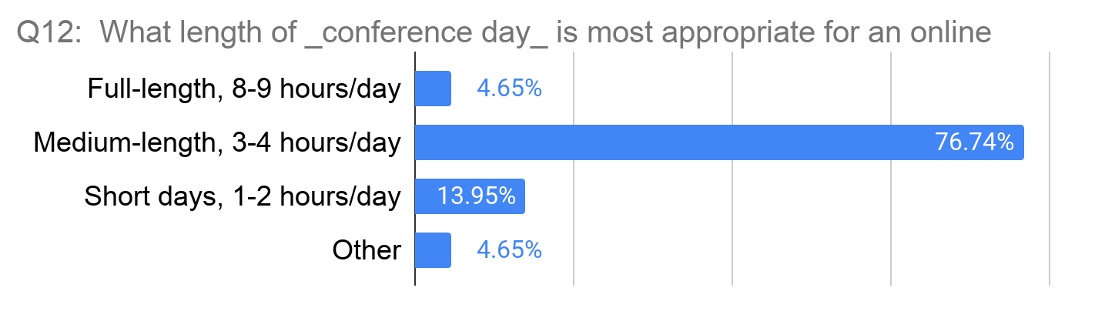}
    \caption{Summary of answers for Q12.}
    \label{fig:Q12}
\end{figure}

Q12~(Figure~\ref{fig:Q12}) indicates over three-quarters of the respondents prefer the medium-length day chosen for ICPE. Some would have liked it even shorter; under 5\% would have liked a full-length day format.

\begin{figure}[h]
    \centering
    \includegraphics[width=\linewidth]{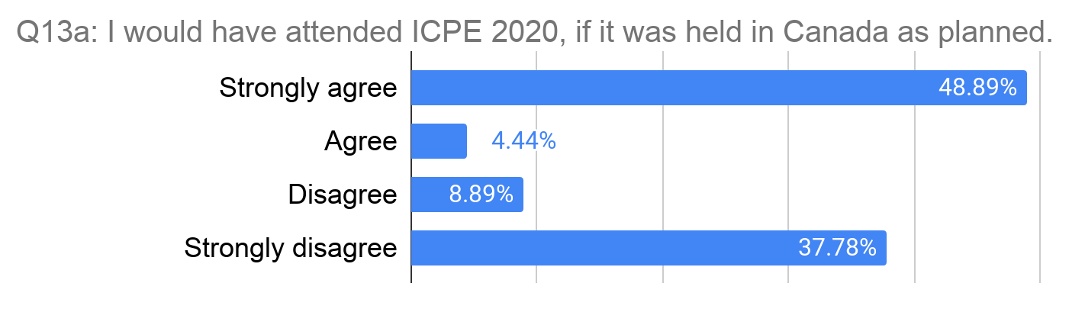}
    \includegraphics[width=\linewidth]{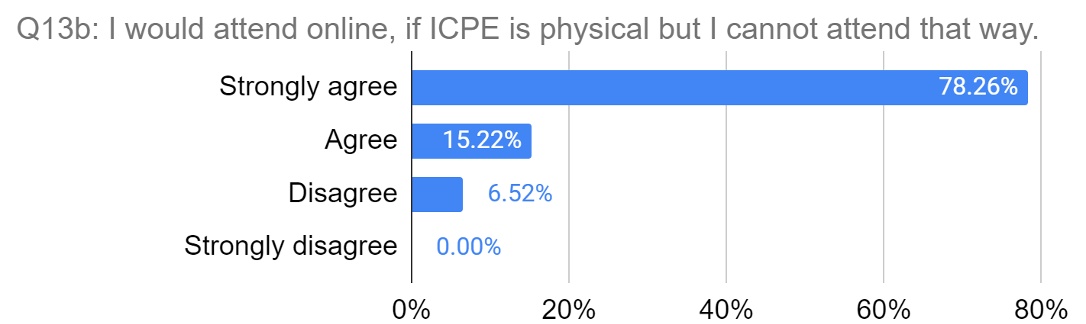}
    \includegraphics[width=\linewidth]{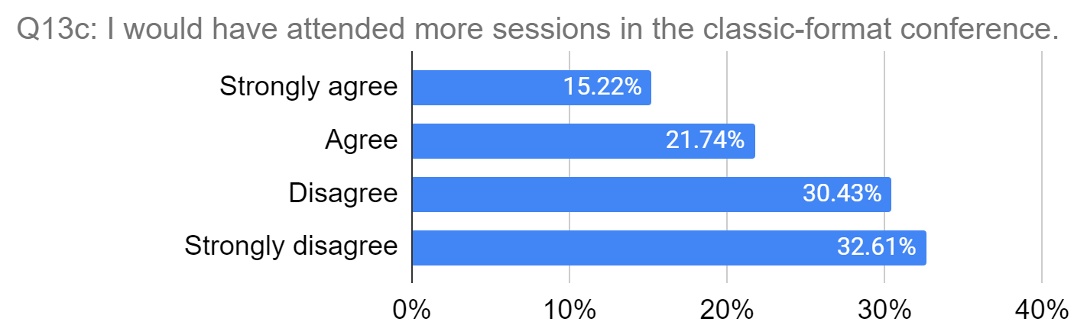}
    \includegraphics[width=\linewidth]{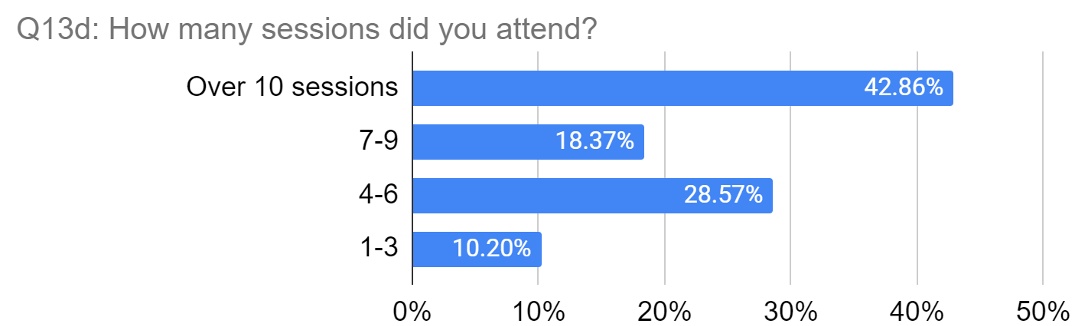}
    \caption{Summary of answers for Q13.}
    \label{fig:Q13}
\end{figure}

From Figure~\ref{fig:Q13}, Q13a and Q13b indicate there are good reasons to organize the conference online: it allows broadening participation.

Q13c gives another reason to allow (also) online attendance: people can attend more sessions. The answers to Q13d show a majority of our respondents have attended at least 7 sessions, with over 40\% attending over 10. We asked more about this: 11 of our 16 sessions were attended by at least 60\% of the respondents. (The informal social sessions were attended by just over a quarter of the attendees, which is less than the typical attendance at a conventional conference.). Further details of the graphical representation of the questions and corresponding responses can be found in the appendix section.

\begin{figure}[h]
    \centering
    \hspace*{-1.0cm}
    \includegraphics[width=1.2\linewidth]{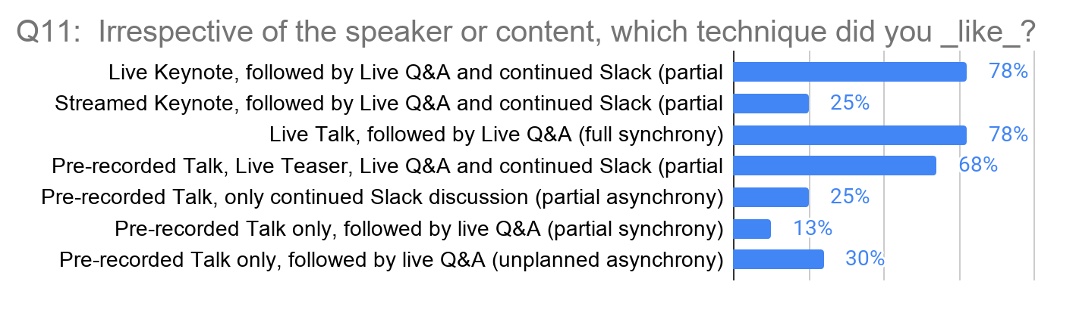}
    \caption{Summary of answers for Q11.}
    \label{fig:Q11}
\end{figure}

From Q11 answers~(Figure~\ref{fig:Q11}), we observe that live talks are preferred over any other form of pre-recorded talk. However, pre-recorded talks that include even a short live teaser are almost as appreciated, but consume only a fraction of the time allocated for the subject and thus leave much more time for Q\&A.


\newpage{}

\section{Conclusion and Our Advice for Conference Organizers}
\label{sec:conclusion}

The ICPE 2020 experience has been rewarding for all involved. As organizers, we learned many lessons from organizing this conference online, including
\begin{enumerate}
    \item The most important factor for the success of a professional meeting is human. Focus on increasing the availability and attention of your audience. Be as flexible as possible. Engage with the community. Also, train the community (humanely).
    
    \item Communication with people is of key relevance. Authors of accepted papers needed some encouragement to prepare slides, videos, the 1-slide pitch, a preprint of their articles, etc. PC members need to be tempted to volunteer in the revision of the material and to actively participate in online sessions.
    
    \item A community that has as alternative not attending the conference at all will tolerate many mishaps and issues with the infrastructure. Under these circumstances, the infrastructure is not a major issue.
    
    \item The infrastructure needed to organize a conference online includes many components. Prioritize flexibility and redundancy in supporting various modes of communication.
    
    \item  Testing the infrastructure is essential, but perhaps not much more so than in conventional conferences.
    
    \item Organizing medium-length conference days of 3-4 hours is perceived as better than using shorter or longer alternatives. 
    
    \item Daily Cafe (open) sessions at the end of each day led to useful discussion, mostly for community building and about how to organize better.
    
    \item One innovation in format: the live pitch of about 2 minutes is an efficient replacement for the live talk, which leaves more time for Q\&A. Consequently, Q\&A can be better than in conventional conferences.
    
    \item The online conference can offer various advantages over the classic physical format; perhaps a mixed mode would become the norm in the future.
    
    \item One main advantage of organizing online: enlarging the audience at a fraction of the cost. Compared with a physical conference, organizing online can lead to lower material costs, lower environmental costs, (often) lower costs for the employer, and (often) lower personal costs. The flexibility of joining or leaving at any moment also decreases the opportunity costs.
    
    \item One main victim of organizing online: the personal connection. Currently, there is no substitute for the physical attendance of a conference talk, or for the advantages in establishing collaborations of physical meetings. Also, the social event still does not have a good equivalent in the digital world. The existing tools, and perhaps also the formats tried so far, simply cannot deliver the same experience.

\end{enumerate}

\section*{Acknowledgment}

The authors would like to acknowledge the sponsors of the ACM/SPEC ICPE 2020 conference. We also want to thank all the organizers. The conference could not have been organized without their combined and generous contribution.

The authors further thank Snigdha Singh, of KIT, Germany, who kindly provided most of the visual analysis presented in the Appendix. This extends the analysis in Section~\ref{sec:feedback} and provides a fuller view of the feedback offered by conference attendees.

\bibliographystyle{plain}
\bibliography{ms}

\begin{thebibliography}{1}

\bibitem{report:acm20}
{ACM Presidential Task Force}.
\newblock Virtual conferences: A guide to best practices.
\newblock Report on What Conferences Can Do to Replace Face-to-Face Meetings,
  \url{https://www.acm.org/virtual-conferences}, Apr 2020.
\newblock Accessed: April 2020.

\bibitem{proc:icpe20ws}
Jos\'{e}~Nelson Amaral, Anne Koziolek, Catia Trubiani, and Alexandru Iosup.
\newblock Icpe 2020 companion, proceedings of workshops, tutorials, and demos.
\newblock ACM conference proceedings
  \url{https://dl.acm.org/doi/proceedings/10.1145/3375555}, Apr 2020.
\newblock Accessed: April 2020.

\bibitem{proc:icpe20}
Jos\'{e}~Nelson Amaral, Anne Koziolek, Catia Trubiani, and Alexandru Iosup.
\newblock Icpe 2020 full proceedings of the main conference.
\newblock ACM conference proceedings
  \url{https://dl.acm.org/doi/proceedings/10.1145/3358960}, Apr 2020.
\newblock Accessed: April 2020.

\bibitem{blog:cacm:Bonifati+20}
Angela Bonifati, Giovanna Guerrini, Carsten Lutz, Wim Martens, Lara Mazilu,
  Norman Paton, {Marcos Antonio} {Vaz Salles}, Marc~H. Scholl, and Yongluan
  Zhou, editors.
\newblock {\em Holding a Conference Online and Live, Due to {COVID-19}}. {ACM},
  Apr 2020.
\newblock Accessed: April 2020.

\bibitem{website:icpe20}
{ICPE Team}.
\newblock Icpe 2020 website.
\newblock \url{https://icpe2020.spec.org/}, Apr 2020.
\newblock Accessed: April 2020.

\bibitem{youtube:icpe20}
{ICPE Team}.
\newblock Icpe 2020 youtube channel.
\newblock
  \url{https://www.youtube.com/channel/UCq_1N-drTe7fdWLPRRbcihQ/videos}, Apr
  2020.
\newblock Accessed: April 2020.

\bibitem{slack:icpe}
{ICPE Team}.
\newblock Icpe slack workspace.
\newblock \url{https://icpeworkspace.slack.com}, Apr 2020.
\newblock Accessed: April 2020.

\bibitem{blog:cacm:LarusCS20}
James~R. Larus, Luis Ceze, and Karin Strauss, editors.
\newblock {\em The {ASPLOS} 2020 Online Conference Experience}. {ACM}, Mar
  2020.
\newblock Accessed: April 2020.

\bibitem{DBLP:conf/ifip/Strachey59}
Christopher Strachey.
\newblock Time sharing in large, fast computers.
\newblock In {\em {Information Processing, Proceedings of the 1st International
  Conference on Information Processing, UNESCO, Paris 15-20 June 1959}}, pages
  336--341, 1959.

\end{thebibliography}



\section*{Author's Biographies}

Prof.dr.ir. {\bf Alexandru Iosup} is full tenured professor and University Research Chair at Vrije Universiteit (VU) Amsterdam, and member of the Young Royal Academy of Arts and Sciences of the Netherlands. He received his PhD in computer science from TU Delft, the Netherlands. He is the chair of the Massivizing Computer Systems research group at the VU and of the SPEC-RG Cloud group. His work in distributed systems and ecosystems has received prestigious recognition, including the 2016 Netherlands ICT Researcher of the Year, the 2015 Netherlands Higher-Education Teacher of the Year, and several SPEC community awards SPECtacular (last in 2017). He served as a Program Chair for ICPE 2020. He can be contacted at \url{A.Iosup@vu.nl} or \url{@AIosup}.\\

Prof.dr. {\bf Anne Koziolek} is a professor at Karlsruhe Institute of Technology (KIT), Germany, since 2019. She received the Diploma degree in informatics from University of Oldenburg, Germany, in 2007, and the PhD degree in informatics from KIT in 2011. She received in 2008 a research fellowship of the German National Academic Foundation ("Studienstiftung des deutschen Volkes”). She conducted research also at the FZI Forschungszentrum Informatik and at the University of Zurich, Switzerland. Her research attracted numerous citations and nominations, including a Test of Time award from ICPE. She served in prestigious roles, such as co-chair, for ICPE, WOSP-C, QRASA, etc. She served as a General Chair for ICPE 2020. \\

Dr. {\bf Catia Trubiani} is Assistant Professor at the Gran Sasso Science Institute (GSSI) Italy, i.e., an international PhD school and a center for research and higher education. She received her PhD in computer science at the University of L’Aquila, and she was a research fellow for a EU FP7 project at the University of Rome. Her work on modeling and analyzing large-scale software systems has received prestigious recognition, including the Best Research Paper Award at ECSA in 2015 and ICPE in 2011, the Microsoft Azure Research Award (market value: 40k USD) in 2014. Among various international projects in which she has been involved, she is principal investigator of the GSSI unit for a MIUR-PRIN project (Young Line action). She served as a Program Chair for ICPE 2020.  She can be contacted at \url{catia.trubiani@gssi.it} or \url{@CatiaTrubiani}.\\

Prof.dr. {\bf Jos\'{e} Nelson Amaral}, a Computing Science professor at the University of Alberta, Ph.D. from The University of Texas at Austin in 2004, has published in optimizing compilers and high-performance computing. Scientific community service includes general chair for the 23rd International Conference on Parallel Architectures and
Compilation Techniques in 2014, for the International Conference on Performance Engineering (ICPE) in 2020, and for the International Conference on Parallel Processing in 2020. Accolades include ACM Distinguished Engineer, IBM Faculty Fellow, IBM Faculty Awards, IBM CAS "Team of the Year", awards for excellence in teaching, and the GSA Award for Excellence in Graduate Student Supervision. He is an elected member of the Board of Directors of the Standard Performance Evaluation Corporation (SPEC), one of the ICPE sponsoring organizations. He served as a General Chair and as Local Chair for ICPE 2020.\\

Dr. {\bf Andre B. Bondi} is a highly experienced consultant, researcher, and author in software performance engineering. He recently  completed a year as an Adjunct Professor of Software Engineering at Stevens Institute of Technology in Hoboken, New Jersey. He has served as a co-chair of the steering committee of ICPE since 2012. He spent over ten years working on performance at AT\&T Labs and its predecessor, AT\&T Bell Labs, before holding senior performance positions in two startups and more than ten years at Siemens Corporate Technology, where he initiated their U.S.-based performance engineering practice. He received the A. A. Michelson Award from the Computer Measurement Group in 2016. He received his Ph.D. in computer science from Purdue University and an M.Sc. in statistics from University College London. He spent a semester as a visiting professor of computer science at the University of L’Aquila in 2016. He served as a member of the ICPE Task Force for Organizing ICPE Online.\\

Dr. {\bf Andreas Brunnert} is founder of RETIT GmbH, a software and consulting company focused on software performance (\url{https://www.retit.de}). Andreas gained more than 15 years' experience in the field of software performance engineering in his various roles at IBM and while leading a research group focused on software performance at an institute of the Technical University of Munich. In recent years, he has focused on the integration of measurement- and model-based performance evaluation approaches. He holds a Diploma in computer science from University of Applied Sciences Brandenburg, Germany, a M.Sc. in information systems from University of Bamberg, Germany, and a Ph.D. in computer science from Technical University of Munich. He served as the Chair of ICPE Industry Track.

\appendix

\section{Analysis (Cont'd.)}

\begin{figure}[h]
    \centering
    \hspace*{-1cm}
    \includegraphics[width=1.2\linewidth]{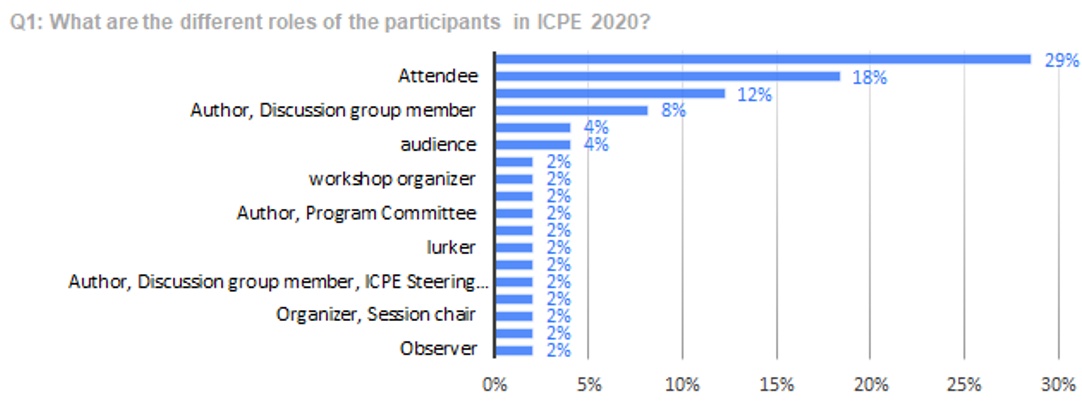}
    \caption{Summary of answers for Q1.}
    \label{fig:Q1}
\end{figure}

The answers to question Q1~(Figure~\ref{fig:Q1}) indicate the answers to our survey were provided primarily by authors of articles presented in the conference (29\%), followed by regular attendees (18\%). This is expected, as the commitment of authors may drive them to further take the time to give feedback, but may bias the results toward the positive, as authors could compare favorably the organization of the conference (even online) against not organizing the conference at all.\\

\begin{figure}[h]
    \centering
    \includegraphics[width=\linewidth]{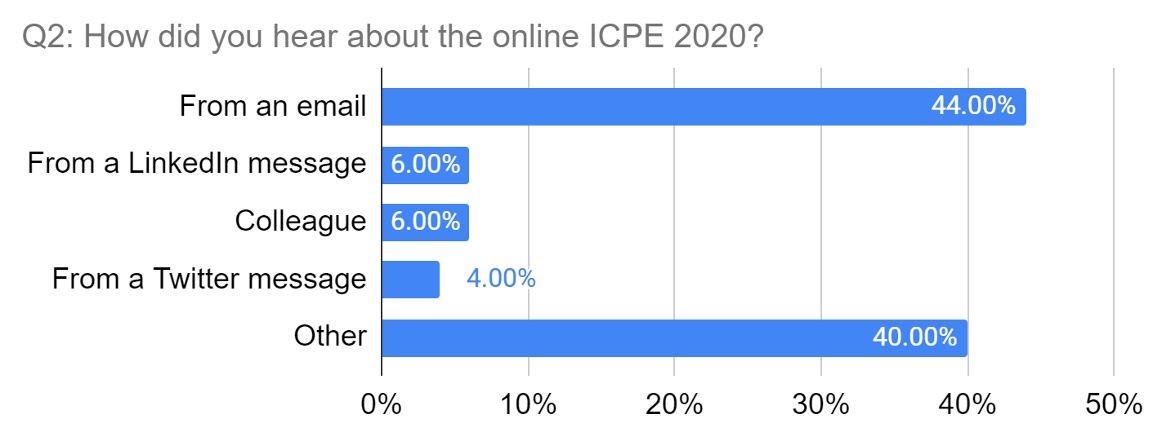}
    \caption{Summary of answers for Q2.}
    \label{fig:Q2}
\end{figure}

The answers to question Q2~(Figure~\ref{fig:Q2}) reveal that the direct-messaging strategy of the conference organizers worked very well. By far the most selected answer (44\% of the respondents) was that attendees learned about ICPE 2020 going online from email. The answers can include multiple choices, but email still appears as the only consistent choice. This may be surprising in an age where social media receives so much attention, but it is an important indication that professional meetings should still use email as one of the main communication channels. \\

\begin{figure}[h]
    \centering
    \includegraphics[width=\linewidth]{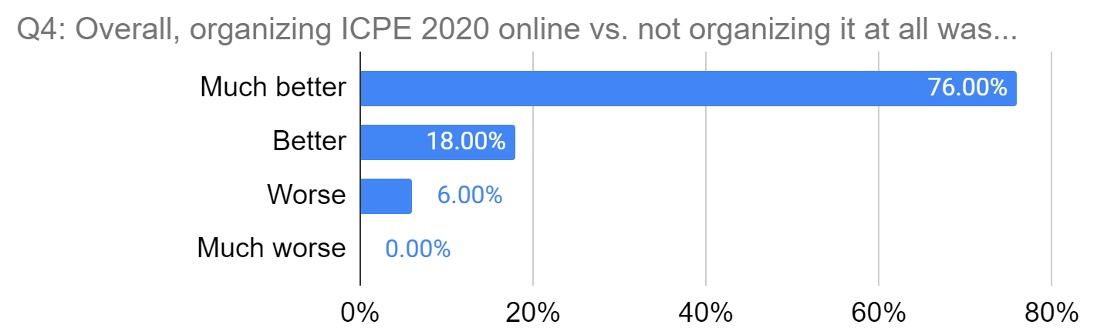}
    \caption{Summary of answers for Q4.}
    \label{fig:Q4}
\end{figure}

The answers for Q4~(Figure~\ref{fig:Q4}) reveal that organizing the conference online was much preferred by the respondents (94\%, of which over three-quarters replied it was much better to do so), over not organizing at all (only 6\%).\\

\begin{figure}[h]
    \centering
    \includegraphics[width=\linewidth]{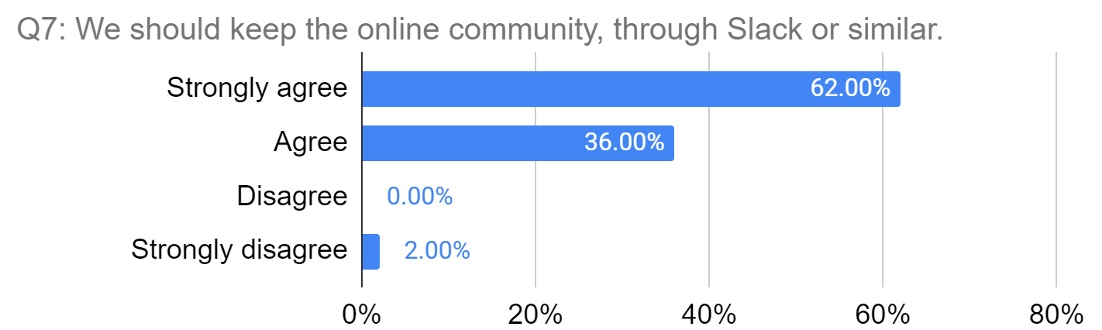}
    \caption{Summary of answers for Q7.}
    \label{fig:Q7}
\end{figure}

The answers to Q7~(Figure~\ref{fig:Q7}) indicate the use of Slack is perceived as positive by a large part of the community. Over 95\% of the respondents indicate Slack should continue to be used as a mechanism to maintain the ICPE community year-round.\\

\begin{figure}[h]
    \centering
    \hspace*{-0.75cm}\includegraphics[width=\linewidth]{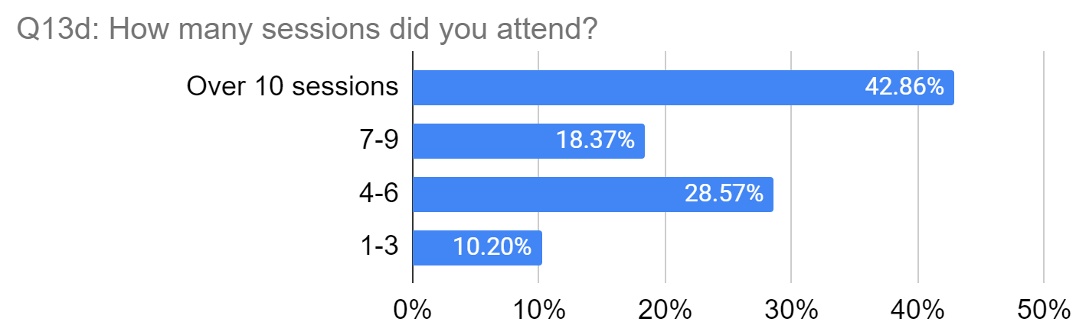}
    \caption{Summary of answers for Q13d.}
    \label{fig:Q13d}
\end{figure}

The answers to Q13d~(Figure~\ref{fig:Q13d}) indicate the use of online technology allowed the ICPE audience to attend a large number of sessions. There were in total 16 sessions. Nearly two-thirds of the attendees attended at least 7 sessions, and over 40\% of the attendees attended over 10 sessions (about two-thirds of the sessions on offer). Based on our experience with conferences in general, this high level of attendance may even exceed that of conventional conferences.\\

\begin{figure}[h]
    \centering
    \hspace*{-0.45cm}\includegraphics[width=\linewidth]{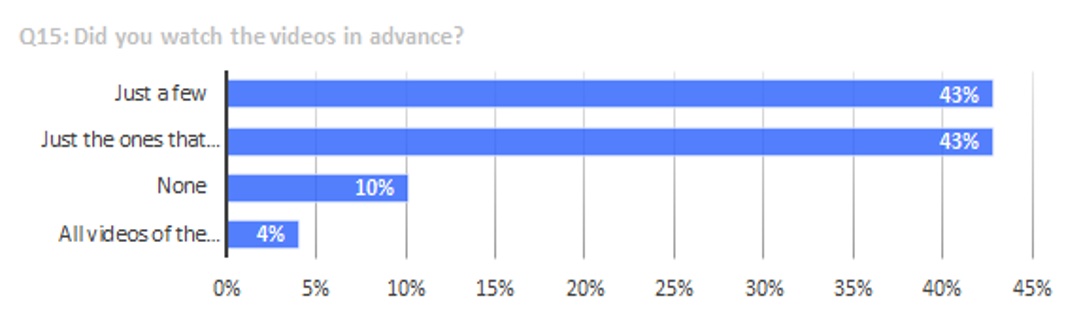}
    \caption{Summary of answers for Q15.}
    \label{fig:Q15}
\end{figure}

Q15~(Figure~\ref{fig:Q15}) reveals that watching videos in advance of the day of the related Q\&A session was not done by over 40\% of the audience. Thus, we derive from this that the idea of the short pitch at the start of each session was meaningful in providing at least a good starting point for the discussion. (In conversations with attendees, we learned some of them watched the videos during the pitches and while the Q\&A was ongoing, using the 2x speedup option of YouTube videos.)\\

\begin{figure}[h]
    \centering
    \includegraphics[width=\linewidth]{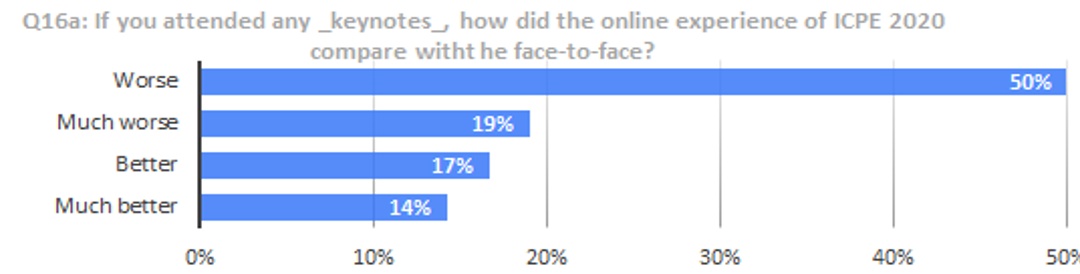}\\
    \hspace*{0.4cm}\includegraphics[width=\linewidth]{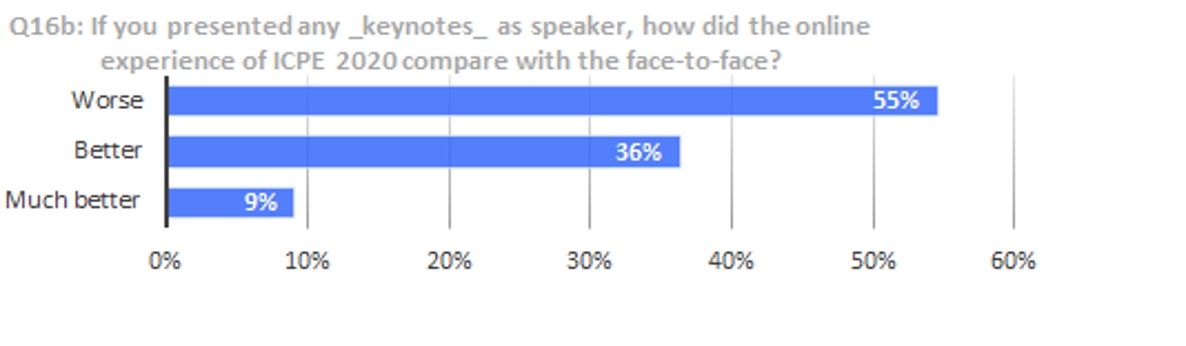}
    \caption{Summary of answers for Q16.}
    \label{fig:Q16}
\end{figure}

Q16a and Q16b~(Figure~\ref{fig:Q16}) reveal that both the audience and the speakers experienced the keynotes worse than they normally would in a conventional conference. (This was not caused by the speakers themselves, who were rated highly by the audience.) This leads us to conclude that, even if the attendees agree that the tools we used for the conference were the best available, these tools still cannot deliver an experience close to live-attendance.\\

\begin{figure}[h]
    \centering
    \hspace*{-0.5cm}\includegraphics[width=\linewidth]{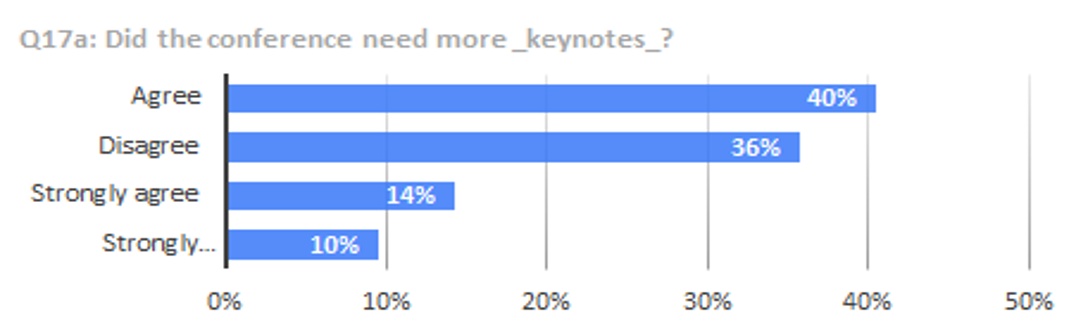}
    \caption{Summary of answers for Q17a.}
    \label{fig:Q17a}
\end{figure}

Q17a~(Figure~\ref{fig:Q17a}) strengthens our conclusion that the attendees appreciated the presence of good content, more than how it was supported by the communication tools. A majority of the attendees (54

However, the opinions are almost split over this issue. With the argument that having a compact program was desirable and useful for ICPE 2020, we conjecture we had the right amount of keynotes.\\

\begin{figure}[h]
    \centering
    \hspace*{-0.5cm}\includegraphics[width=1.2\linewidth]{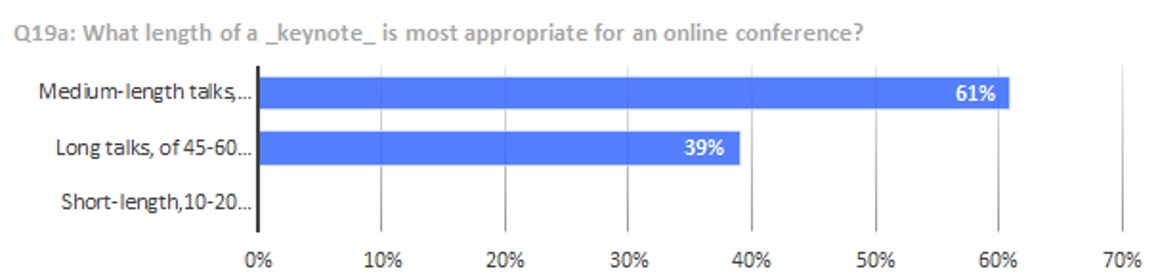}\\
    \hspace*{-0.35cm}\includegraphics[width=1.2\linewidth]{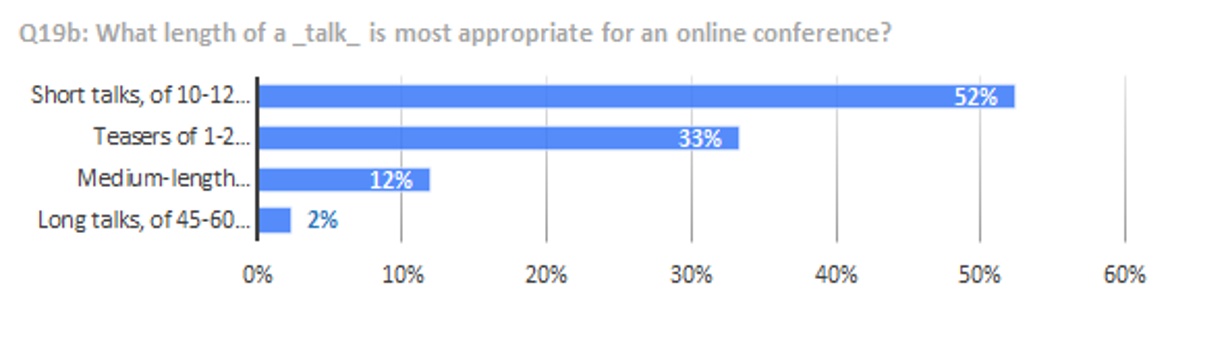}\\
    \vspace*{-0.8cm}
    \caption{Summary of answers for Q19.}
    \label{fig:Q19}
    \vspace*{-0.5cm}
\end{figure}

From Q19a and Q19b~(Figure~\ref{fig:Q19}), we observe that shorter is in general preferable to our attendees. One important decision we took, the trade-off of not organizing even short talks but to switch entirely to teasers, vs. how much the audience would prefer, leads to an interesting conclusion. The attendees would have preferred short talks, but, as indicated by the experiment conducted by one of the workshops, would not have attended them because the conference day would have run too long into the night (for European attendees). This trade-off, of talk-length (and thus day-length) vs. attendance, remains at the core of organizing conferences, whether they are organized virtually or physically.\\

\begin{figure}[h]
    \centering
    \hspace*{-1.5cm}
    \includegraphics[width=1.2\linewidth]{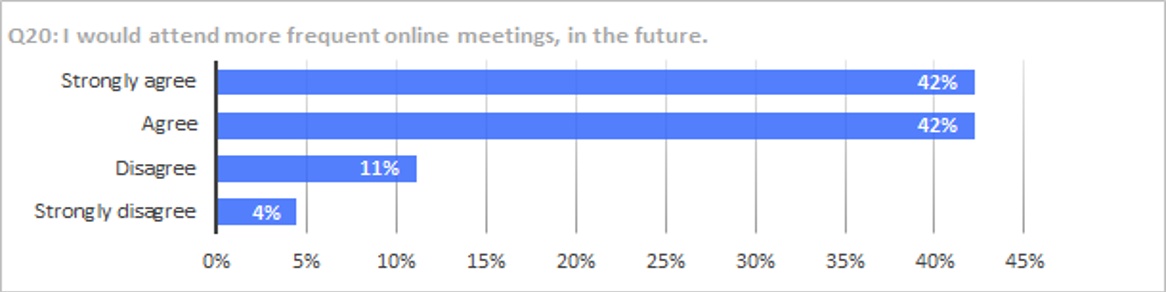}
    \vspace*{-0.8cm}
    \caption{Summary of answers for Q20.}
    \label{fig:Q20}
    \vspace*{-0.5cm}
\end{figure}

The answers to Q20~(Figure~\ref{fig:Q20}) indicate the attendees would agree to online meetings being held more frequently. This underlines a strength of online conferences: they can allow the community to meet more frequently, albeit, at the cost of some of the quality. A combination of several online conferences and one main (yearly) conference seems a good idea to explore in the future.\\

\begin{figure}[h]
    \centering
    \hspace*{-0.5cm}
    \includegraphics[width=1.2\linewidth]{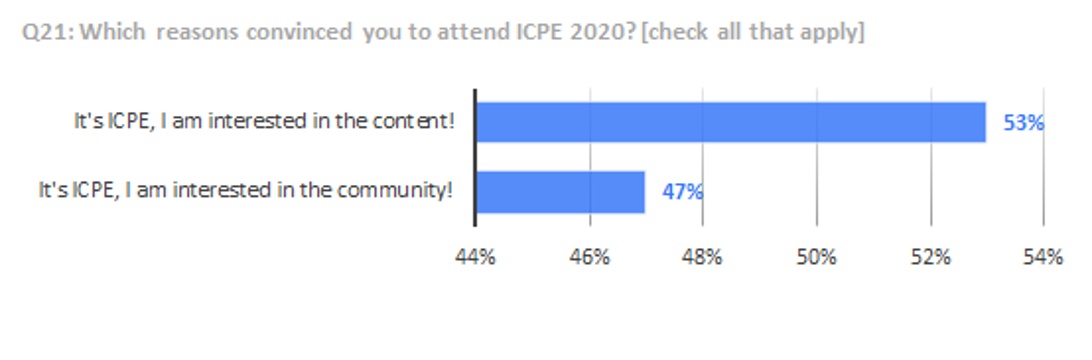}
    \vspace*{-1cm}
    \caption{Summary of answers for Q21.}
    \label{fig:Q21}
    \vspace*{-0.25cm}
\end{figure}

The answers to Q21~(Figure~\ref{fig:Q21}) are consistent with the idea that the ICPE audience was very interested in both the content (over 90\%) and the community (over 80\%). This emphasizes the need to find the right tools for community engagement.\\

\begin{figure}[h]
    \centering
    \hspace*{-0.2cm}
    \includegraphics[width=1.2\linewidth]{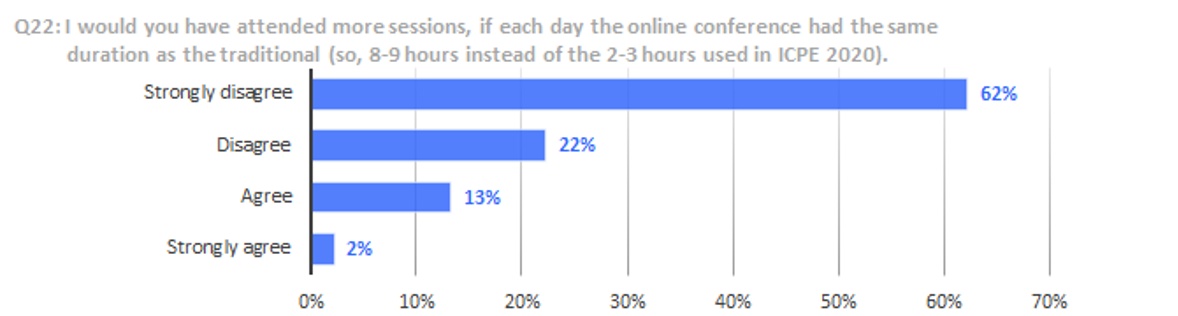}
    \vspace*{-0.75cm}
    \caption{Summary of answers for Q22.}
    \label{fig:Q22}
    \vspace*{0.5cm}
\end{figure}

The answers to Q22~(Figure~\ref{fig:Q22}) indicate that, should we have used a longer conference-day, and in particular the length of a conventional conference, the audience would not have attended more sessions. We conclude our design choice regarding the trade-off day-length vs. attendance was appropriate. The presence of the trade-off and this conclusion need to be communicated more clearly to the audience, to prevent confusion -- as exemplified by some of the answers to Q19a and Q19b.\\

\begin{figure}[h]
    \centering
    \hspace*{0.3cm}\includegraphics[width=\linewidth]{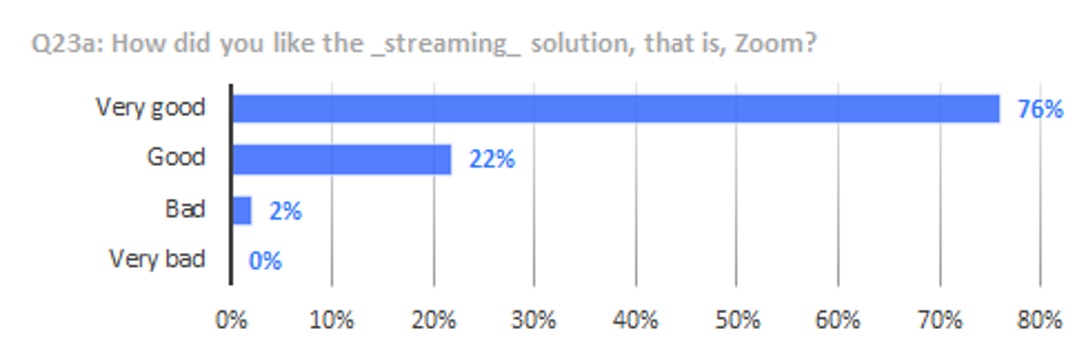}\\
    \hspace*{-1.5cm}\includegraphics[width=1.25\linewidth]{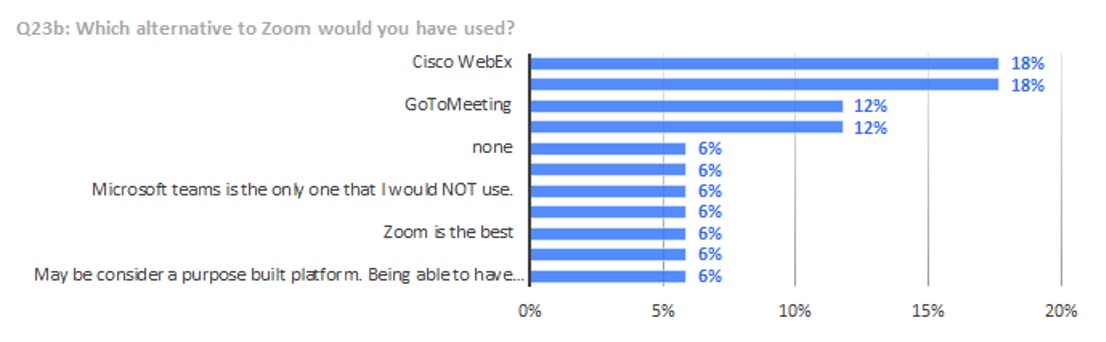}
    \vspace*{-1cm}
    \caption{Summary of answers for Q23.}
    \label{fig:Q23}
\end{figure}

Figure~\ref{fig:Q23} summarizes the answers related to the streaming solution, Zoom. Almost all of the respondents to Q23a (98\%) agreed that Zoom is a good or very good solution for streaming. Yet, this does not make it sufficient to emulate the physical meeting, as indicated by the answers given to Q16a and Q16b.

The answers to Q23b are not surprising: there exist many streaming tools that people have experience with and like. Among the most suggested are GoToMeeting, Cisco WebEx, Microsoft Teams / Link, and Google Meet / Hangout. We, the authors and other organizers, have conducted experiments using GoToMeeting, and had personal experience with the other three leading suggestions. \\

\begin{figure}[h]
    \centering
    \hspace*{0.25cm}\includegraphics[width=\linewidth]{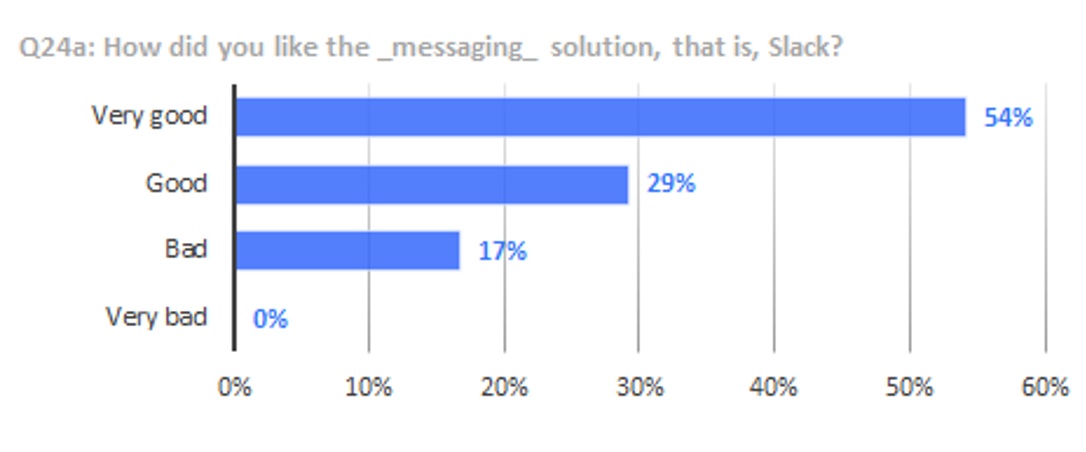}\\
    \includegraphics[width=\linewidth]{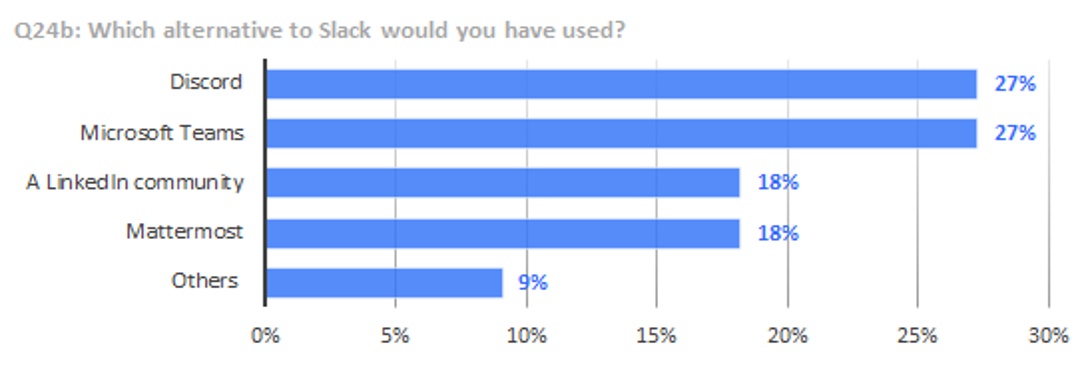}\\
    \hspace*{0.6cm}\includegraphics[width=\linewidth]{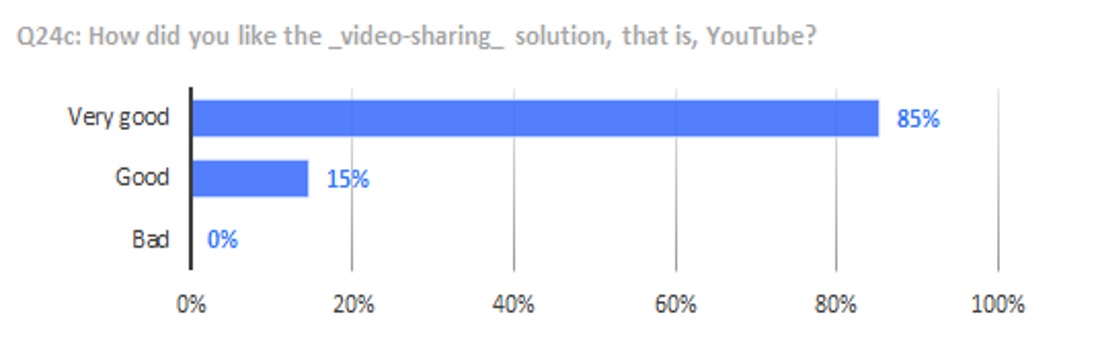}\\
    \hspace*{-0.3cm}\includegraphics[width=\linewidth]{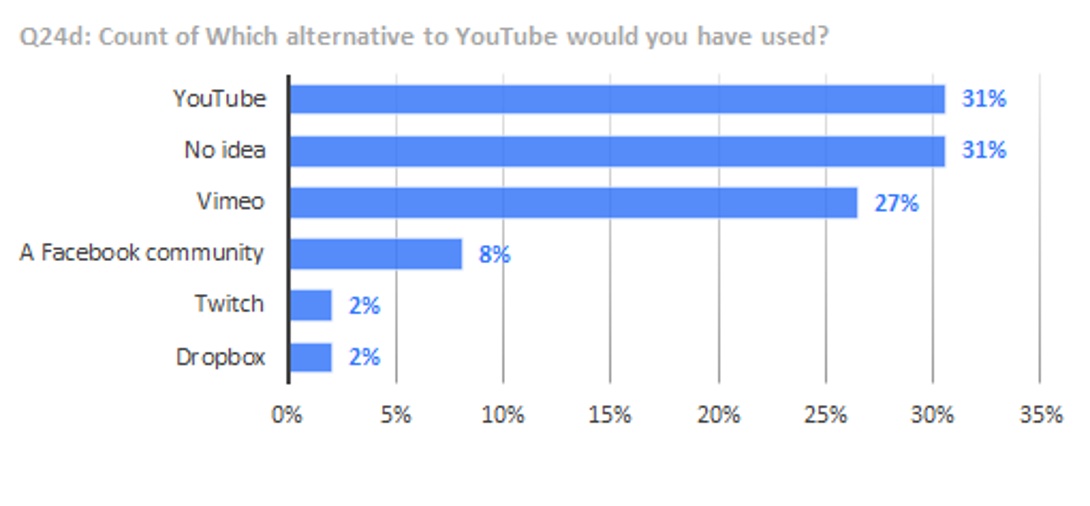}\\
    \vspace*{-0.8cm}
    \caption{Summary of answers for Q24.}
    \label{fig:Q24}
    \vspace*{-0.25cm}
\end{figure}

Figure~\ref{fig:Q24} summarizes the answers related to the messaging solution, Slack, and for the video-sharing solution, YouTube. 
For Q24a, respondents indicate their acceptance of Slack as a messaging solution, with over 80\% of them agreeing it is at least good. However, relatively to Zoom for streaming, the approval of Slack as very good for messaging is much lower. From the open-form feedback, we observe some of the attendees found Slack as too complex when starting with it, a situation also identified by the organizers of ASPLOS. It may be good for Slack to provide a simpler interface for starting users, with more advanced features enabled from specialized menus. 

Similarly to the answers given for Zoom as a streaming platform, the attendees gave as answers for Q24b a variety of other choices. Microsoft Teams, Discord, using a LinkedIn community, and Mattermost (an open-source alternative very similar to Slack) were the most popular choices. Members of the team had extensive experience with the last three. This indicates future ICPE events could experiment with using Microsoft Teams instead of Slack.

Similarly to Zoom, the respondents to Q24c see YouTube as good or very good. The answer most given indicates YouTube is very good at what it can provide in this area.
Last, and perhaps confusingly, over 60\% of the attendees gave as an alternative to YouTube ``no idea'' or YouTube itself (presumably, as an indication there is no reasonable alternative for this context).\\

\begin{figure}[h]
    \centering
    \vspace*{-0.5cm}
    \hspace*{-1cm}
    \includegraphics[width=1.2\linewidth]{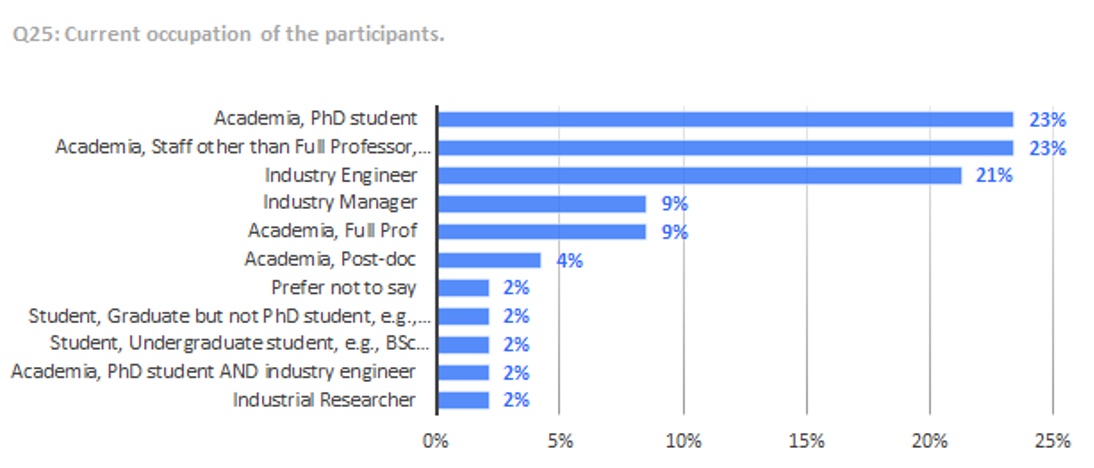}\\
    \vspace*{-0.25cm}
    \caption{Summary of answers for Q25.}
    \label{fig:Q25}
    \vspace*{-0.3cm}
\end{figure}

The answers to Q25~(Figure~\ref{fig:Q25}) indicate the diversity of our audience, occupation-wise. We observe a balanced participation of industry staff, academic staff, and students. \\

\vspace*{1cm}

\begin{figure}[h]
    \centering
    \hspace*{0.1cm}\includegraphics[width=0.7\linewidth]{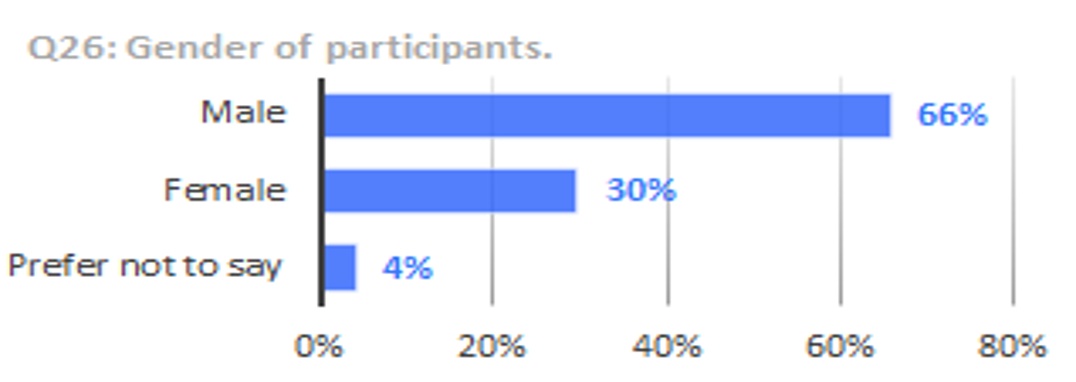}
    \caption{Summary of answers for Q26.}
    \label{fig:Q26}
\end{figure}

The answers to Q26~(Figure~\ref{fig:Q26}) reveal the gender participation in the conference. In our experience, the fraction of attendees identifying as female is among the highest in the field.\\

\begin{figure}[h]
    \centering
    \hspace*{-0.5cm}
    \hspace*{-0.3cm}\includegraphics[width=1.2\linewidth]{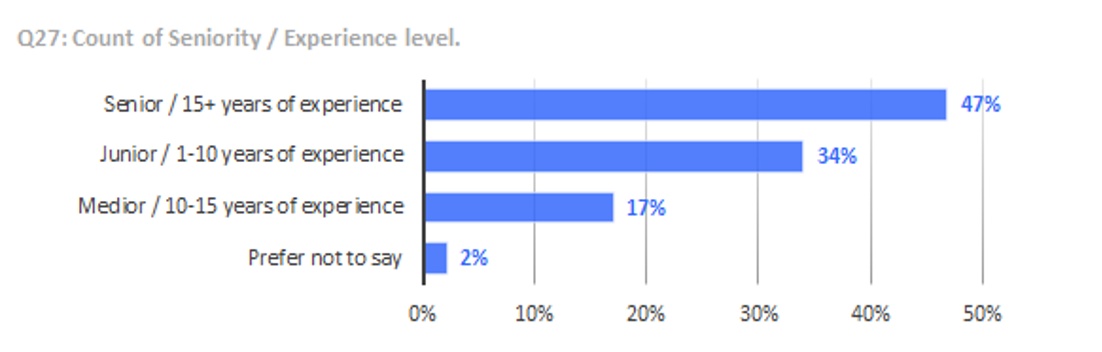}
    \caption{Summary of answers for Q27.}
    \label{fig:Q27}
\end{figure}

The results for Q27~(Figure~\ref{fig:Q27}) are consistent with a diverse community, but also with the idea that the reputation of ICPE is one of experience-sharing.\\

\begin{figure}[h]
    \centering
    \includegraphics[width=\linewidth]{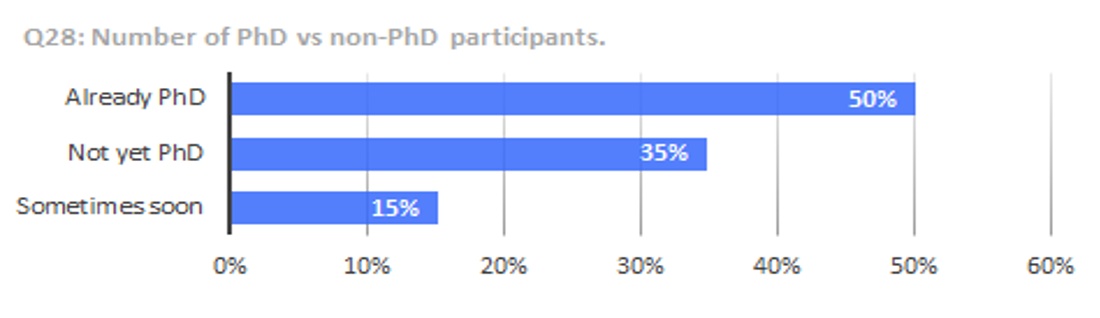}
    \caption{Summary of answers for Q28.}
    \label{fig:Q28}
\end{figure}

A large fraction of the participants have already or expect soon to have a PhD degree, as indicated by the answers received for Q28~(Figure~\ref{fig:Q28}).\\

\begin{figure}[h]
    \centering
    \hspace*{-0.9cm}
    \includegraphics[width=1.25\linewidth]{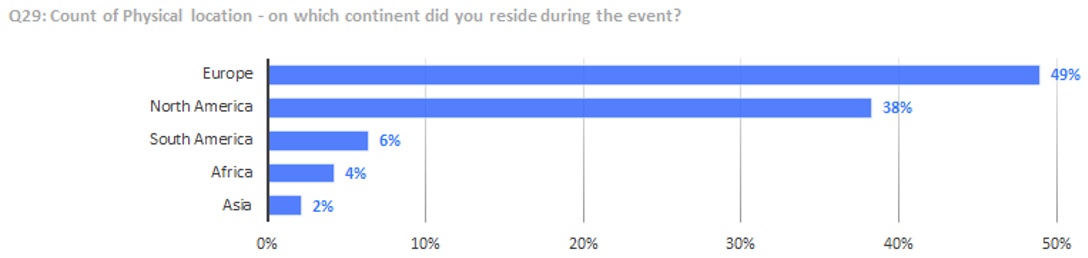}
    \vspace*{-0.75cm}
    \caption{Summary of answers for Q29.}
    \label{fig:Q29}
    \vspace*{-0.5cm}
\end{figure}

The answers to Q29~(Figure~\ref{fig:Q29}) indicate strong participation from Europe and North America, but also low participation from other continents. Asia, which has given a large fraction of participants during the previous edition organized in India, is not well-represented this time. This indicates one of the major drawbacks of online conferences: their operational hours can make it difficult to participate from entire continents. (The alternative offered by conventional conferences requires attendees from these continents to travel to the location of the conference.)

\begin{figure}[h]
    \centering
    \hspace*{-0.4cm}
    \includegraphics[width=1.2\linewidth]{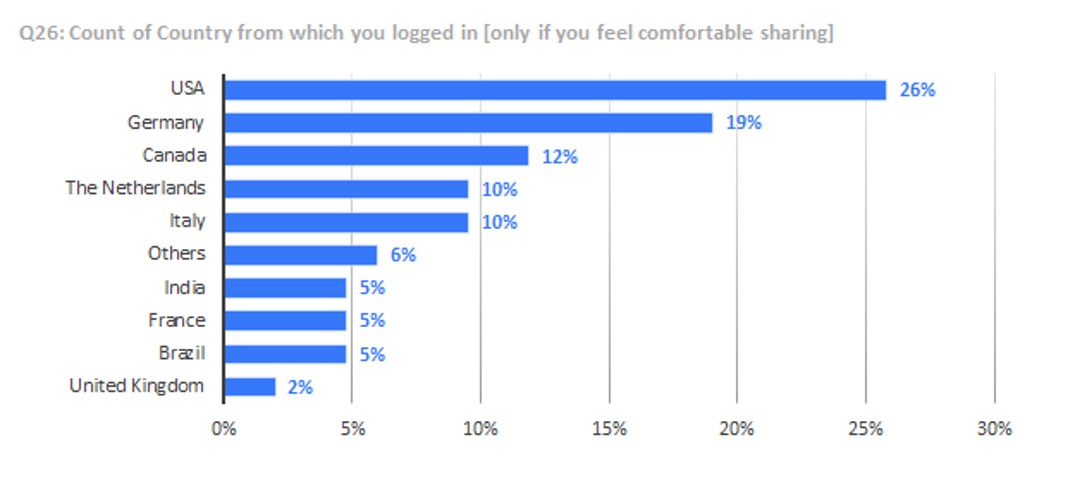}
    \vspace*{-1cm}
    \caption{Summary of answers for Q30.}
    \label{fig:Q30}
\end{figure}

The answers to Q30~(Figure~\ref{fig:Q30}) indicate participation from many countries, but also that countries like Germany, USA, Canada, Italy, and the Netherlands provide a majority of the participants. We see it as the task of the Steering Committee to consider if broadening and balancing participation across more countries is possible. 

\end{document}